\documentclass[preprint]{aastex}
\begin{document}

\title{A Correction for IUE UV Flux Distributions from Comparisons with 
CALSPEC}

\author{Ralph~C.\ Bohlin\altaffilmark{1}, and Luciana Bianchi\altaffilmark{2}}
\altaffiltext{1}{Space Telescope Science Institute, 3700 San Martin Drive,
Baltimore,  MD 21218, USA}
\altaffiltext{2}{Department of Physics and Astronomy, The Johns Hopkins
University, Baltimore, MD 21218, USA}

\begin{abstract}  

A collection of spectral energy distributions (SEDs) is available in the
\emph{Hubble Space Telescope} (HST) CALSPEC database \textbf{that is} based on 
calculated model atmospheres for pure hydrogen white dwarfs (WDs). A much larger
set ($\sim$100,000) \textbf{of} UV SEDs covering the range (1150-3350~\AA) with
somewhat lower quality are available in the IUE database. IUE low-dispersion
flux distributions are compared with CALSPEC to provide a correction that places
IUE fluxes on the CALSPEC scale. While IUE observations are repeatable to only
4--10\% in regions of good sensitivity, the average flux corrections have a
precision of  2--3\%. Our re-calibration places the IUE flux scale on the
current UV reference standard and is relevant for any project based on IUE
archival data, including our planned comparison of GALEX to the corrected IUE
fluxes. \textbf{IUE SEDs may be used to plan observations and cross-calibrate
data from future missions, so the IUE flux calibration must be consistent with
HST instrumental calibrations to the best possible precision.}\end{abstract}


\section{Introduction}

Flux values in physical units for astronomical objects are required to make
comparisons to physical models, for example measuring relative fluxes of
redshifted supernovae Ia's and, thus, in determining the nature of the dark
energy that is driving the observed accelerating cosmic expansion
\citep{scolnic14}. A good set of primary and secondary absolute flux standards
is in the
CALSPEC\footnote{http://www.stsci.edu/hst/observatory/crds/calspec.html}
archive, where the flux calibration relies on pure hydrogen white dwarf (WD)
models for G191B2B, GD153, and GD71 to calibrate HST/STIS spectrophotometry and
to establish UV/optical/near-IR flux standards \citep{bohlin14}.

The IUE spacecraft was launched 1978 January 26 and was sensitive in each of two
UV channels that cover the far-UV (FUV, 1150--2000~\AA) and the near-UV
(NUV,ß1850-3350~\AA) \citep{boggess1978}. Each spectroscopic channel had a
low-resolution and an echelle high-resolution mode with a prime and a redundant
camera for each channel, i.e. SWP and SWR in the FUV and LWP and LWR in the NUV.
Because of better photometric repeatability and higher sensitivity, the low
dispersion modes of SWP, LWP, and LWR with resolution R$\sim$6~\AA\ are of
interest here. The short-wavelength redundant camera SWR was not used for many
science observations. Each camera could obtain spectra with
either a large aperture (L) of $\sim$10x20\arcsec\ or a small aperture (S) of 
$\sim$3\arcsec\ diameter. The L-aperture data are photometrically repeatable,
while the S-aperture transmits 20--70\% of the incident light but does provide
useful  spectral energy distributions (SEDs) when normalized to the L-aperture.

The IUE database represents a homogenous collection of 105,584 spectra, 56227 in
the Short-Wavelegth (SW) range (1150-2000~\AA) and 49357 in the Long Wavelength
(LW, 1850$-$3350~\AA)\footnote{The numbers are for the total of the archival
datasets, although some spectra are for calibration purposes, such as "WAVECAL"
or "FLOOD"}. Although surpassed in quality by UV spectral data from HST
instruments such as FOS, HRS, and STIS \citep{bohlin01}, the IUE spectral
collections remain a frequently used resource with extensive sets of
spectroscopic UV extinction curves and the first reference atlases of UV spectra
of stars \textbf{\citep{wu1996} and galaxies \citep{kinney1996}}. One of our
projects (in preparation) exploits the IUE database to revisit the calibration
of the GALEX spectra by comparing the overlapping target sample. \textbf{GALEX
spectra have a similar wavelength range, $\sim$1350-2830~\AA, and slightly
lower resolution and S/N than the IUE low-dispersion modes.} However, the GALEX
spectral database constitutes a fainter UV spectral collection  than the IUE
set. Because the IUE faint limits are close to the GALEX bright flux safety
limits, there is a small overlap of the two datasets, which can be used to place
the two benchmark spectral collections onto a consistent flux scale. 

Section 2 compares the HST/STIS UV spectrophotometry to the set of stars in
common with IUE to establish the corrections needed to bring IUE fluxes onto the
CALSPEC scale and to estimate the precision of IUE. Section 3 describes the
problematic cases, while Section 4 details the electronic access to our results.
In preparation are companion papers that compare CALSPEC and IUE SEDs with a set
of matched GALEX spectral SEDs and that present the entire database of GALEX
grism flux distributions.

\section{Comparision of IUE and CALSPEC} % Section 2

The IUE data analyzed here are extracted from the Mikulski Archive for Space
Telescopes (MAST)\footnote{http://archive.stsci.edu/iue/}. This Archive contains
both the original IUE data products from the IUESIPS processing system
\citep{bohlin1996} and from the NEWSIPS uniform re-processing of the entire
archive \citep{nichols1996}. The NEWSIPS version provides a somewhat more
photometric product in most cases and uses the optimal extraction algorithm of
\citet{kinney1991}. Unless otherwise explicitly stated, all discussion of IUE
data refers to the \textbf{*.MXLO files of }NEWSIPS SEDs.

\begin{deluxetable}{llcccc}     %Table1
\tablewidth{0pt}
\tablecolumns{6}
\tablecaption{Low-Dispersion IUE Data Used for Comparing IUE and 
	CALSPEC SEDs}
\tablehead{
\colhead{Star} &V (mag) &\colhead{CALSPEC Name} &\colhead{SWP} &\colhead{LWP} 
				&\colhead{LWR}}
\startdata
 G191B2B     &11.78  &g191b2b\_mod\_010.fits       &56 &39 &10\\
 GD153       &13.35  &gd153\_mod\_010.fits         &10 &8 &2  \\
 GD71        &13.03  &gd71\_mod\_010.fits          &10 &8 &1  \\
GRW+70$^{\circ}$5824  &12.77  &grw\_70d5824\_stisnic\_007   &10 &9 &1\\
 HZ21        &14.69  &hz21\_stis\_004.fits	   &17 &9 &4\\
 HZ43        &12.91  &hz43\_stis\_004.fits	   &12 &1 &7\\
% LDS749B &lds749b\_stisnic\_006.fits   &4 &2 &2\\
\enddata
\label{table:calspec}
\end{deluxetable}

Six WD stars with multiple IUE spectra and with complete CALSPEC UV coverage of
the IUE spectral range of 1150--3350~\AA\ are selected for comparison with their
NEWSIPS SEDs, as summarized in Table~\ref{table:calspec}, where the number of
good IUE observations appear in the final columns for each IUE camera. The
comparisons of our co-adds of the good, low-resolution IUE data in
Table~\ref{table:calspec} with the tabulated CALSPEC SEDs define corrections to
the NEWSIPS flux scale. The models, rather than the observations, are the
reference SEDs for the primary flux standards G191B2B, GD153, and GD71 that set
the HST/STIS flux scale. As part of the co-adding process, each IUE observation
is examined for abnormalities, which includes such problems as an exposure too
short to produce reasonable signal-to-noise (S/N), a low or zero signal from a
miscentered target acquistion, or solar contamination at the longest wavelengths
that happened  occasionally near the end of the mission. There are three such
problem cases for G191B2B (SWP: 15727, 48543, 55661) and six for HZ21 (SWP:
21845, 21846, 21847, LWP: 02465, 02466, 02472), where these bad observations are
not included in the Table~\ref{table:calspec} counts. The data points that are
flagged as bad data quality, such as saturation and reseau marks, are omitted,
while the remaining good data points are averaged using the exposure times as
weights.  The reseau and saturation flagging is somewhat inadequate, so the
flags are expanded by one point in each direction. At wavelengths where there
are no remaining valid points in the individual co-adds, the assumed flux is a
linear interpolation across the gap.

\begin{figure}			%fig1
\centering 
\includegraphics*[height=6.5in]{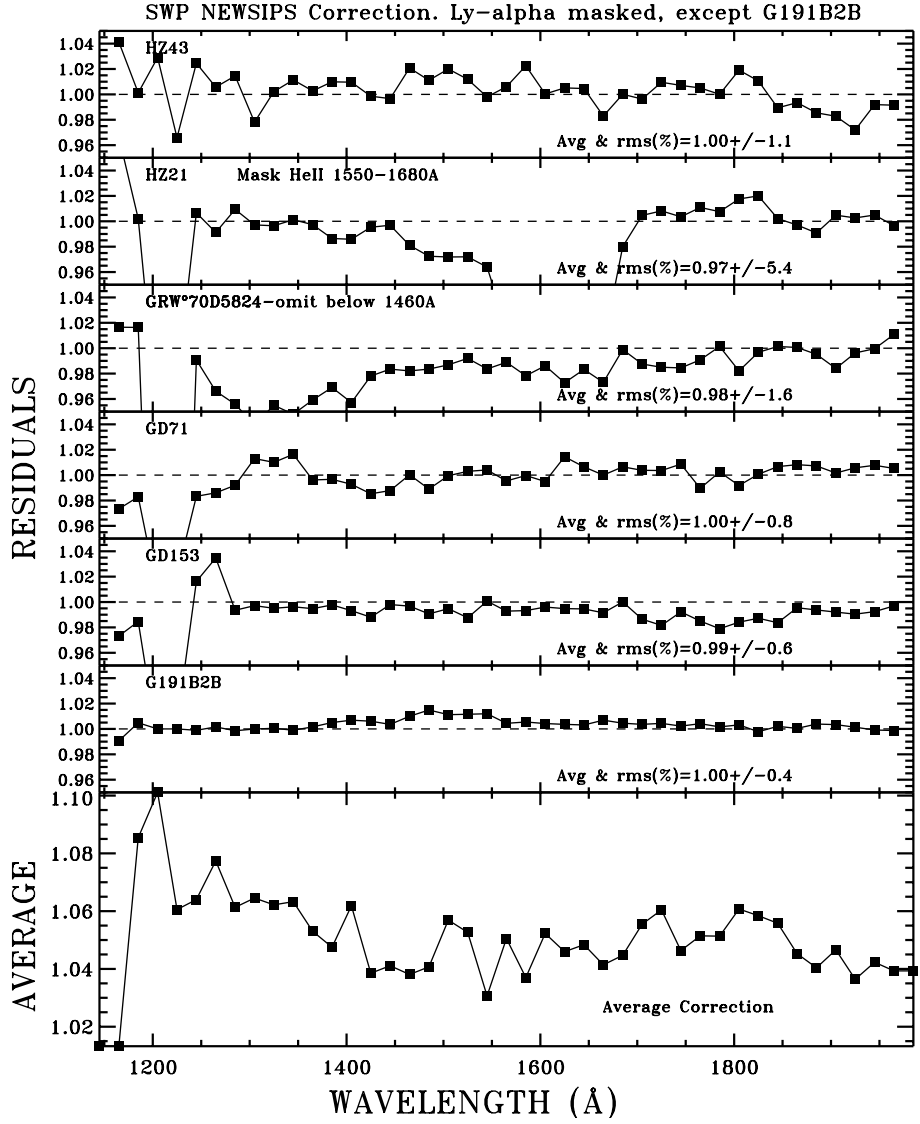}
\caption{\baselineskip=12pt
Average ratio for the IUE SWP camera of CALSPEC/NEWSIPS fluxes at the bottom and
residual ratios for each star, i.e. the residuals show the difference between
the correction derived for each star and the average correction ratio. IUE
NEWSIPS fluxes should be multiplied by the average corrections of
Table~\ref{table:corr} to be on the HST/CALSPEC scale. The average and percent
rms deviation of the  residuals are written for each star for the range of
1300--1900~\AA.  \label{swp}} \end{figure}

\begin{figure}			%fig2
\centering 
\includegraphics*[height=6.5in]{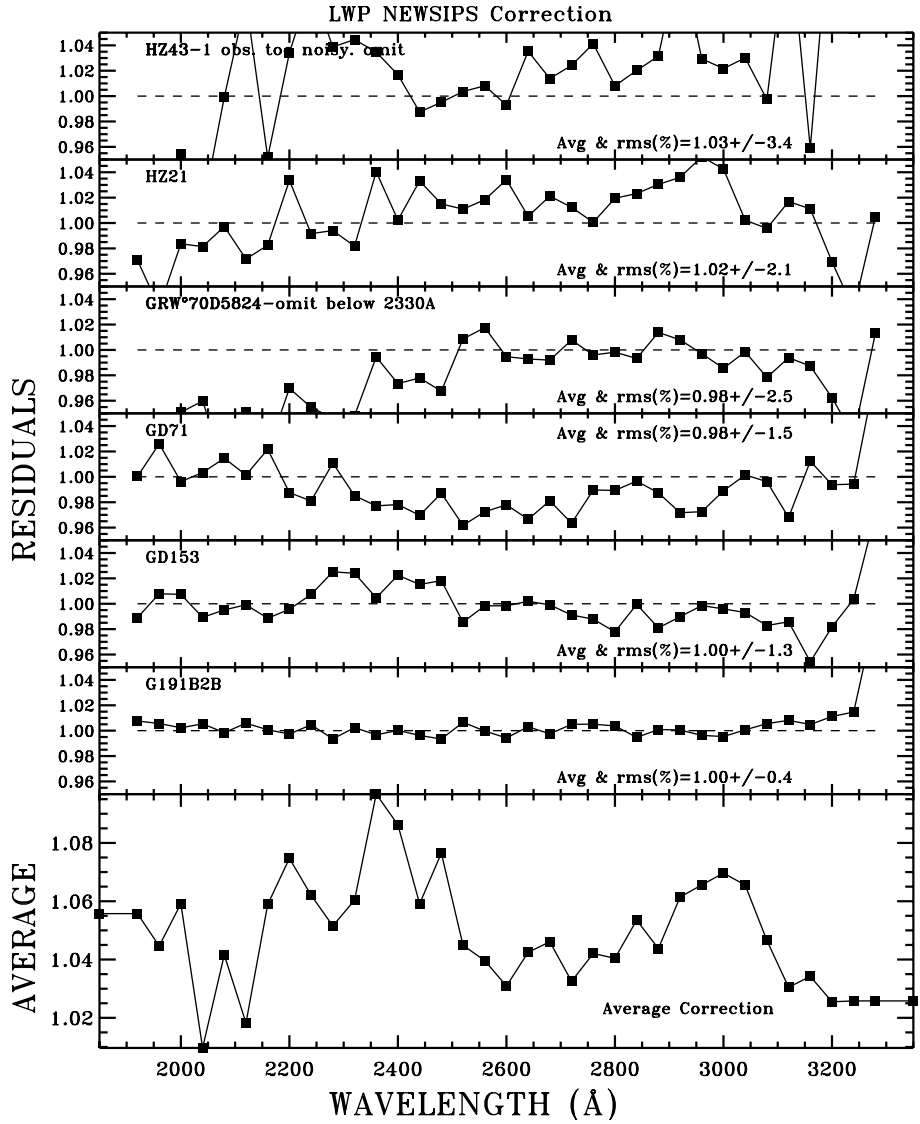}
\caption{\baselineskip=12pt
Average ratio for the IUE LWP camera of CALSPEC/NEWSIPS fluxes at the bottom and
residual ratios for each star for the range of 2100--3000~\AA. 
\label{lwp}} \end{figure}

\begin{figure}			%fig3
\centering 
\includegraphics*[height=6.5in]{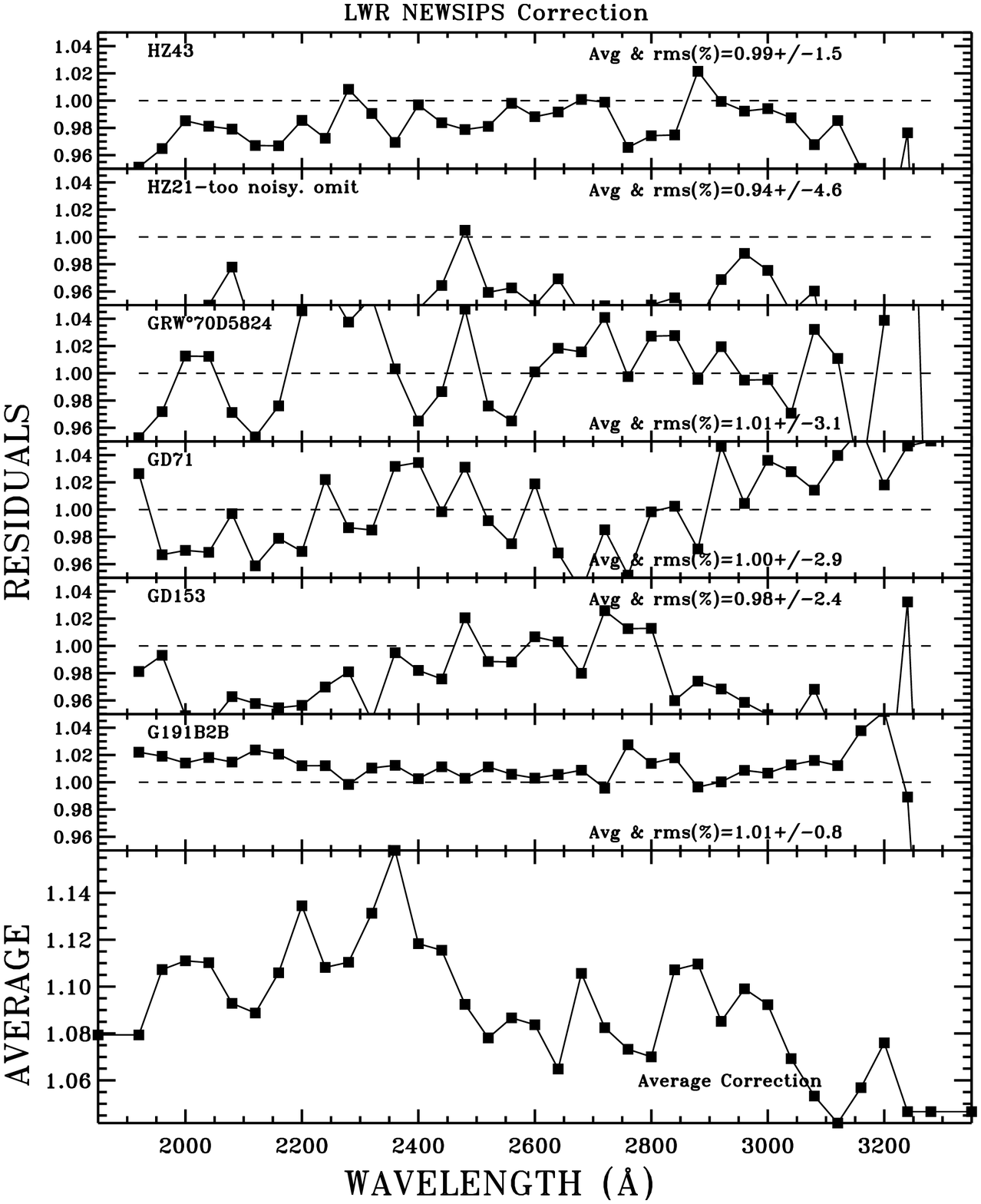}
\caption{\baselineskip=12pt
Average ratio for the IUE LWR camera of CALSPEC/NEWSIPS fluxes at the bottom and
residual ratios for each star for the range of 2100--3000~\AA. 
Results for the faintest star HZ21 are often
off scale at more than 5\% low.
\label{lwr}} \end{figure}

The co-added NEWSIPS fluxes are compared with each CALSPEC star in 20~\AA\ bins
for SWP and 40~\AA\ for LWP and LWR to make individual ratios, and then these
individual ratios are averaged to derive the global average corrections, as
shown in the lowest panels of Figures~\ref{swp}--\ref{lwr}. The global averages
utilize the individual average ratios as weighted by total signal for each star,
so the frequently observed G191B2B dominates the results. Also shown in each
Figure are the individual residuals for the six stars, i.e. the ratio of the
correction derived for each star to the average correction. The average and rms
scatter of these residuals appears on the Figures for the SWP range of
1300--1900~\AA\ and 2100--3000~\AA\ for LWP and LWR.

Table~\ref{table:corr} contains the final global average of corrections from the
six stars, where the regions omitted from these averages are indicated on the
plots. The faintest star HZ21 with only four noisy spectra is omitted entirely
from the LWR co-add. For the well observed stars, i.e. those with seven or more
observations, the separate comparisons with CALSPEC tend to agree fairly well.
However, HZ43 is omitted from the SWP average, because the CALSPEC SED uses the
slightly less precise FOS data below 1700~\AA. The global results in
Table~\ref{table:corr} are extended by one point at each endpoint by  the same
endpoint bin-value in order to achieve complete wavelength coverage.

\begin{deluxetable}{ccccc}     %Table2 - from oldvsnew and its output flxcor.newsips*
% ng \tiny \scriptsize	\footnotesize	\small
\tabletypesize{\scriptsize}
\tablewidth{0pt}
\tablecolumns{5}
\tablecaption{Multiplicative Correction for 
	IUE NEWSIPS Fluxes}
\tablehead{
\colhead{Wavelength~\AA} &SWP Correction &Wavelength~\AA &LWP Correction 
&LWR Correction }
\startdata
 1145.0 &1.013     &1850.0  &1.056     &1.079 \\
 1165.0 &1.013     &1920.0  &1.056     &1.079 \\
 1185.0 &1.085     &1960.0  &1.045     &1.107 \\
 1205.0 &1.101     &2000.0  &1.059     &1.111 \\
 1225.0 &1.061     &2040.0  &1.010     &1.110 \\
 1245.0 &1.064     &2080.0  &1.042     &1.093 \\
 1265.0 &1.077     &2120.0  &1.018     &1.089 \\
 1285.0 &1.061     &2160.0  &1.059     &1.106 \\
 1305.0 &1.065     &2200.0  &1.075     &1.134 \\
 1325.0 &1.062     &2240.0  &1.062     &1.108 \\
 1345.0 &1.063     &2280.0  &1.052     &1.110 \\
 1365.0 &1.053     &2320.0  &1.060     &1.131 \\
 1385.0 &1.048     &2360.0  &1.097     &1.158 \\
 1405.0 &1.062     &2400.0  &1.086     &1.118 \\
 1425.0 &1.039     &2440.0  &1.059     &1.115 \\
 1445.0 &1.041     &2480.0  &1.077     &1.092 \\
 1465.0 &1.038     &2520.0  &1.045     &1.078 \\
 1485.0 &1.041     &2560.0  &1.040     &1.087 \\
 1505.0 &1.057     &2600.0  &1.031     &1.084 \\
 1525.0 &1.053     &2640.0  &1.042     &1.065 \\
 1545.0 &1.031     &2680.0  &1.046     &1.106 \\
 1565.0 &1.051     &2720.0  &1.033     &1.082 \\
 1585.0 &1.037     &2760.0  &1.042     &1.073 \\
 1605.0 &1.052     &2800.0  &1.040     &1.070 \\
 1625.0 &1.046     &2840.0  &1.054     &1.107 \\
 1645.0 &1.048     &2880.0  &1.044     &1.110 \\
 1665.0 &1.041     &2920.0  &1.061     &1.085 \\
 1685.0 &1.045     &2960.0  &1.066     &1.099 \\
 1705.0 &1.055     &3000.0  &1.070     &1.092 \\
 1725.0 &1.060     &3040.0  &1.066     &1.069 \\
 1745.0 &1.046     &3080.0  &1.047     &1.053 \\
 1765.0 &1.051     &3120.0  &1.030     &1.042 \\
 1785.0 &1.051     &3160.0  &1.034     &1.057 \\
 1805.0 &1.061     &3200.0  &1.025     &1.076 \\
 1825.0 &1.058     &3240.0  &1.026     &1.047 \\
 1845.0 &1.056     &3280.0  &1.026     &1.047 \\
 1865.0 &1.045     &3350.0  &1.026     &1.047 \\
 1885.0 &1.040     &	    &	       &      \\
 1905.0 &1.047     &	    &	       &      \\
 1925.0 &1.036     &	    &	       &      \\
 1945.0 &1.042     &	    &	       &      \\
 1965.0 &1.039     &	    &	       &      \\
 1985.0 &1.039     &	    &	       &      \\
\tablecomments{Table 2 is published in its entirety in the machine-readable
format.}
\enddata
\label{table:corr}
\end{deluxetable}

Figure~\ref{swp} shows the best case of SWP, where all six stars are well
observed, and all six residuals agree within an uncertainty of  $\sim$2\% in the
region of higher sensitivity at  1300--1900~\AA. Near Ly-$\alpha$, the spectra
of stars fainter than G191B2B are badly contaminated by geo-coronal Ly-$\alpha$
emission, so the correction near Ly-$\alpha$ is defined by just the one star.
Shortward of Ly-$\alpha$, the IUE sensitivity is poor. The \textbf{omitted}
anomalous regions for  GRW+70$^{\circ}$5824 and HZ21 are discussed in the next
section.

For LWP in the region of higher sensitivity at 2100--3000~\AA, five stars are
well-observed, and only the faintest star HZ21 has residuals bigger than 4\%
in Figure~\ref{lwp}. A
few GD71 residuals approach 4\%; but for the most part, the brighter stars with
the best S/N agree better than our conservative estimate of 3\% accuracy for the
LWP correction. See the next section for a discussion of the anomalous short
wavelength region of the GRW+70$^{\circ}$5824 spectrum.

For LWR, the data are sparse and only HZ43 with seven co-added spectra provides
a worthwhile comparison to G191B2B. Figure~\ref{lwr} shows residuals for HZ43 of
less than 3\% at  2100--3000~\AA, which is our estimate for the precision of the
LWR correction  in Table~\ref{table:corr}.

\begin{figure}			%fig4 
\centering 
\includegraphics*[height=7.5in]{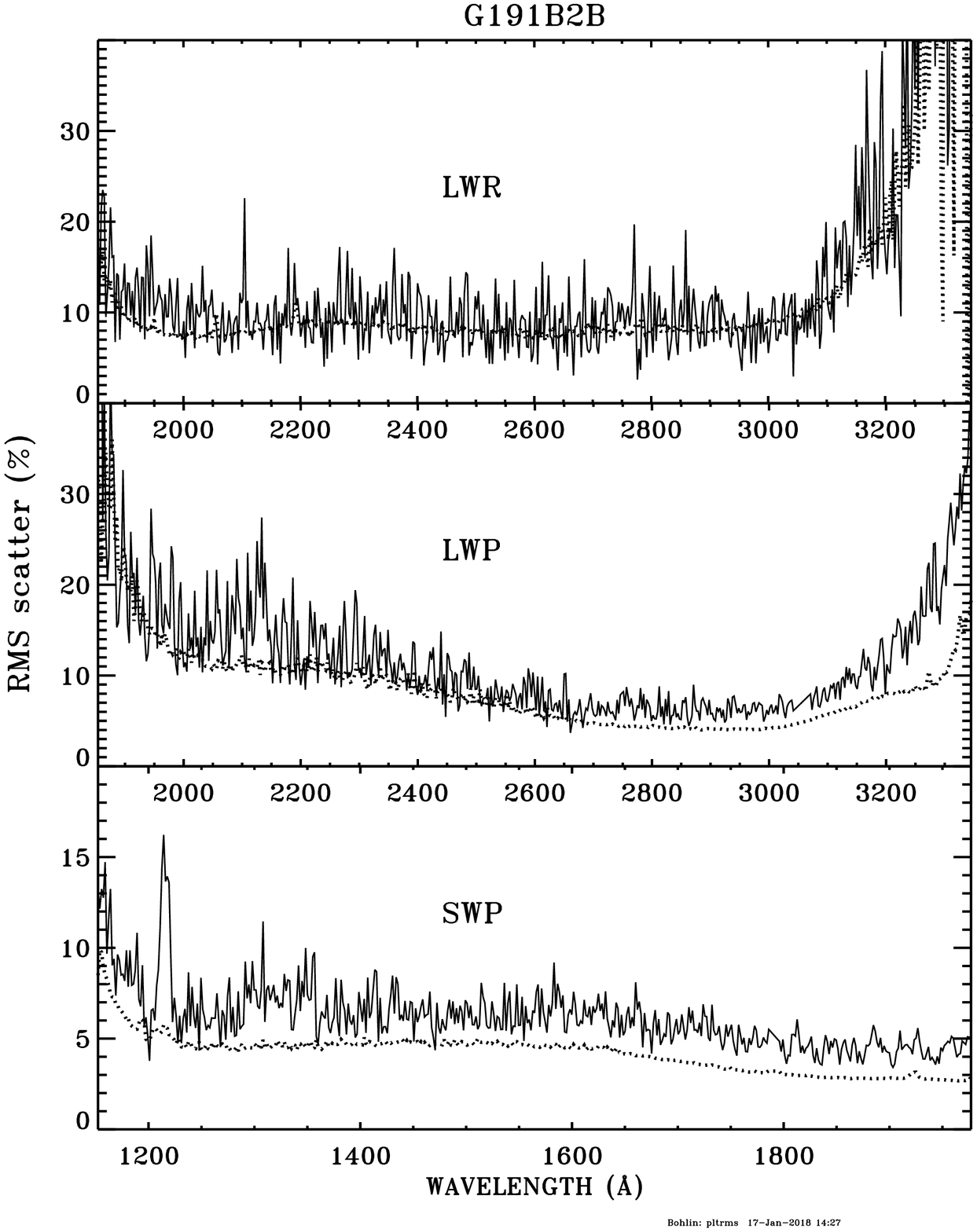}
\caption{\baselineskip=12pt
Precision of a typical IUE spectrum of G191B2B for each camera, as measured by
the rms scatter among the co-added spectra (solid line). The dotted lines are 
the expected rms scatter as computed from the uncertainty in the flux that is
provided in each IUE archival file. These results are for each sample point and
would improve with broad band binning. \label{pltrms}} \end{figure}

\begin{figure}			%fig5
\centering 
\includegraphics*[height=7.5in]{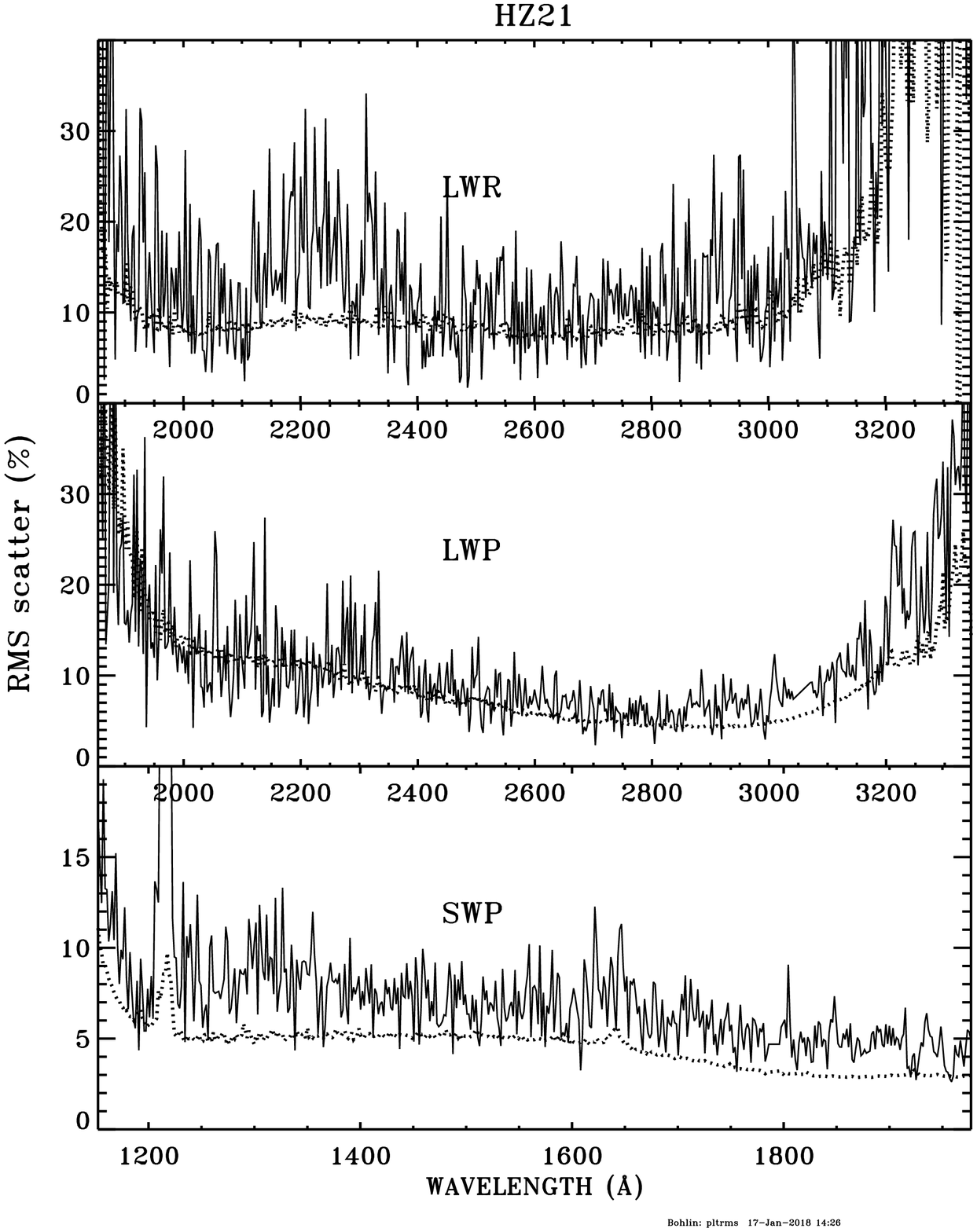}
\caption{\baselineskip=12pt
Precision of IUE spectra of HZ21 for each camera, as in Figure~\ref{pltrms}.
\label{plthz21}} \end{figure}

For individual IUE spectra, the repeatability is a function of the exposure
level and the background noise, both of which vary with wavelength. The
background increases with exposure time and depends on environmental variables,
such as the trapped particle flux in the Van Allen radiation belts. However, the
rms scatter among the co-added spectra provides a measure of the precision of a
typical IUE observation. Figure~\ref{pltrms} shows the rms scatter among the
G191B2B observations of Table~\ref{table:calspec} as the solid lines. The SWP
repeatability ranges mostly from 4--7\%, except at Ly-$\alpha$ and below. The
extra scatter at line-center is due to the variability of the contaminating
geo-coronal Ly-$\alpha$ emission. LWP repeats best at $\sim$6\% from 2600 to
3000~\AA\ with a loss of precision both shortward and longward. LWR repeats at
8--10\% shortward of $\sim$3000~\AA\ and loses precision even faster than LWP
toward longer wavelengths. For our faintest star, HZ21, Figure~\ref{plthz21}
shows a moderate degradation of precision, where the biggest deviation from the
G191B2B results approach a factor of two worse at the shorter wavelengths of
LWR; however, there are only four spectra that define the LWR result. The
NEWSIPS data files contain estimates for the flux uncertainty, which is
propagated to an error-in-the-mean in our co-added and merged files. \textbf{The
IUE NEWSIPS error estimate (sigma) is based on the noise model in the optimal
extraction algorithm. The noise model was developed before 1991 and might not
be a good representation for later observations. Undoubtedly, there are
sources of error in addition to the noise model.} The dotted lines in
Figures~\ref{pltrms}--\ref{plthz21} are the error-in-the-mean multiplied by the
square root of the number of IUE spectra in  the co-adds to compare with the
solid-line rms scatter. The two uncertainty estimates agree fairly well but with
a tendency for the observed rms to be higher, especially for SWP.

\section{Anomalies} % Section 3

\subsection{H21 1640~\AA\ He~II}

\begin{figure}			%fig6
\centering 
\includegraphics*[height=6.5in]{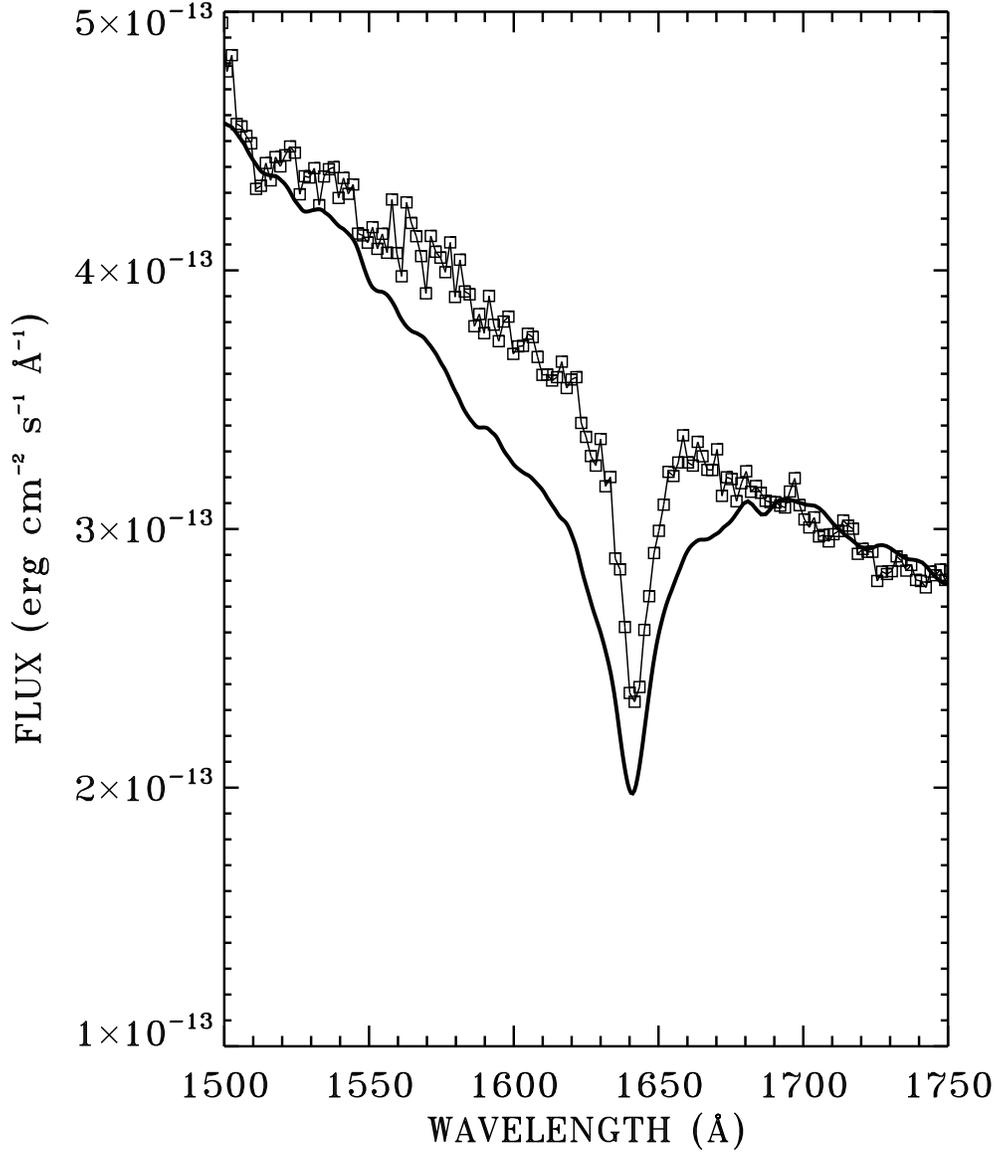}
\caption{\baselineskip=12pt
Discrepant region of HZ21 near the strong He~II 1640.5~\AA\ line. The squares
are the average IUE flux from 1978-1987, while the heavy line is the STIS 
SED from observation o40901nqm in 1997 as convolved to the 6~\AA\ resolution of
IUE.
\label{he2}} \end{figure}

Apparently, the strong, broad He~II line at 1640.5~\AA\ in HZ21 has changed, as
illustrated in Figure~\ref{he2}. The 17 IUE spectra all agree within the
expectations of Figure~\ref{plthz21} and were obtained
in the 1978 April--1987 March epoch, while the CALSPEC/STIS data are from 1997
May. The discrepant region from 1550--1680~\AA\ is excluded from the average
correction factors. The IUE and STIS disagree
by up to $\sim$20\% near the 1640.5~\AA\ line center. 

\subsection{GRW+70$^{\circ}$5824}

\begin{figure}			%fig7 grw.pro
\centering 
\includegraphics*[height=6.5in]{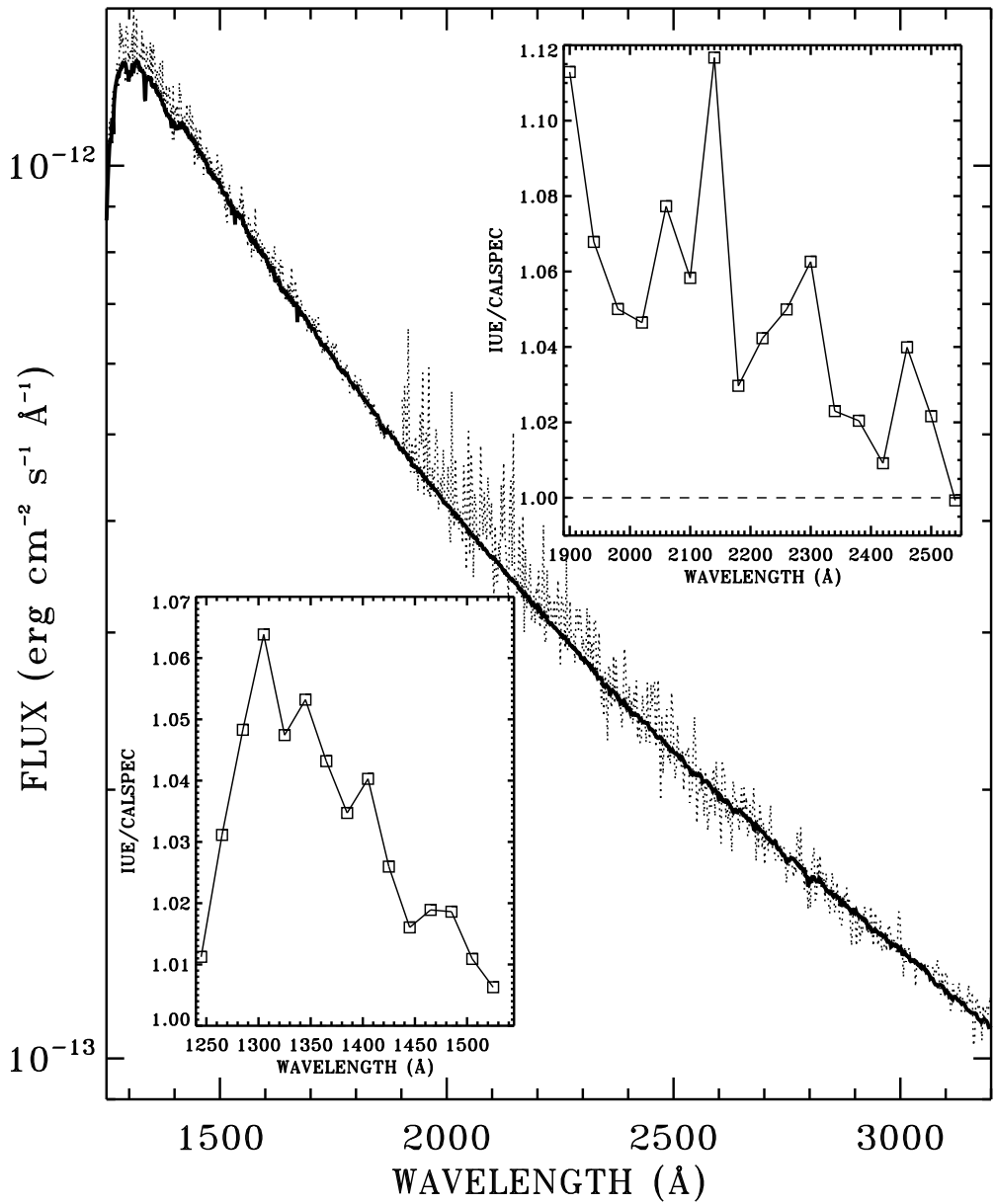}
\caption{\baselineskip=12pt
Discrepant regions of GRW+70$^{\circ}$5824 at the shorter wavelengths of SWP and
LWP. The dotted lines are the average IUE fluxes for SWP and LWP, while the
heavy line is the STIS SED. The inserts show the ratio of the IUE to
CALSPEC/STIS in 20~\AA\ bins for SWP and 40~\AA\ bins for LWP in the regions of
systematic deviation.
\label{grw}} \end{figure}

Both SWP and LWP show discrepancies with CALSPEC/STIS at the shorter
wavelengths, as shown in Figure~\ref{grw}. The IUE SWP flux is systematically
high in the 1250--1500~\AA\ region by up to $\sim$5\%, while LWP is noisy but
is in even worse agreement with STIS below 2500~\AA. However, the single LWR
IUE observation shows no discrepancy. An astrophysical cause of the SWP
and LWP anomalies is hard to imagine in terms of plausible stellar variability.

The STIS data for GRW+70$^{\circ}$5824 are from the sensitivity monitoring
program with $\sim$90 separate STIS observations and constitute one of the most
precise CALSPEC SEDs. Thus, the anomalous fluxes ratios in the insets of
Figure~\ref{grw} are most likely due to IUE instrumental problems. The LWP flux
below 2500~\AA\ shows excess noise, suggesting weak exposure levels in
comparison to the background level; and the increasing error toward shorter
wavelengths is a signature of a slightly low background subtraction that has a
larger effect as the signal drops toward shorter wavelengths. Indeed, the
typical background rate per second is more than double that of GD153, while
the stellar fluxes are comparable. At 2100~\AA, the GRW+70$^{\circ}$5824
flux that is 6\% too high could be caused by only a $\sim$20\% underestimate of
the higher than normal background signal.

The anomalous SWP flux for GRW+70$^{\circ}$5824 is the worst at 1300--1350~\AA\
near the peak of the SWP sensitivity, but our average flux is determined by
unsaturated exposures where the background has little effect on the peak net
signal. Perhaps, the $\sim$5\% flux errors are just a 2.5 sigma anomaly, or
maybe our 2\% uncertainty for the SWP correction is a bit too optimistic. The
original IUESIPS fluxes agree with the NEWSIPS averages shown in
Figure~\ref{grw}.

%\newpage

\section{Electronically Available Results} % Section 4

\textbf{The co-added and merged IUE spectra for the six stars of
Table~\ref{table:calspec} and the Table~\ref{table:corr} NEWSIPS correction
factors are available in the on-line version of this paper and also appear in
MAST as High-Level Science Products 
(HLSP)\footnote{\url{https://archive.stsci.edu/prepds/iue-fluxcal/}}.} The
merged ascii files combine all useful datasets as described in Section 2 and
cross from SWP to the long wavelength (LW) cameras at 1975~\AA. These merged
files contain eight columns: (1) wavelength in \AA, (2) the average net signal
in linearized IUE Flux Number (FN) units per second, (3) the average flux in erg
cm$^{-2}$ s$^{-1}$ \AA$^{-1}$,  (4) the  background signal in the same units as
the net, (5) the formal propagated uncertainty as the error-in-the-mean in flux
units, (6) the number of observations averaged, (7) the total exposure time in
seconds, and (8) the rms scatter among the observations in percent. As our
planned papers appear, more combined IUE spectra of matched sources having GALEX
spectra and the whole collection of GALEX spectra (over 100,000) will also be
\textbf{electronically available. A complete summary of our work along with
the machine readable files also
appears on our 
website\footnote{\url{http://dolomiti.pha.jhu.edu/uvsky/iue-fluxcal/}}. }

\textbf{The DOI for the individual IUE spectra from Table~\ref{table:calspec}
is
\dataset[doi:10.17909/T94M4K]{https://dx.doi.org/10.17909/T94M4K}, while the DOI
for the Tables~\ref{table:corr}--\ref{table:hz43} is
\dataset[doi:10.17909/T9QD6J]{https://dx.doi.org/10.17909/T9QD6J}.
}

\section{Summary}

\textbf{Our comparison of IUE NEWSIPS UV data to the HST-based absolute flux
standards in CALSPEC produces a correction factor for the IUE flux scale. The
corrections are wavelength-dependent and cover the IUE range of sensitivity from
1150 to 3350~\AA. The correction of IUE to CALSPEC ranges from factors of
1.01--1.16. IUE-NEWSIPS *.MXLO fluxes should be multiplied by the
Table~\ref{table:corr} correction factors to be on the HST/CALSPEC flux scale.
This new correction will allow more accurate photometric calibration of existing
UV datasets, such as GALEX through a comparison with corrected IUE NEWSIPS
fluxes, as well as providing critical support for calibration and observation
planning of future space missions.}

\acknowledgments

Karen Levy, Bernie Shiao, and Randy Thomson provided many helpful suggestions
and clarifications with regard to the IUE MAST database. Scott Fleming entered
our HLSP into MAST. Support for this work
was provided by NASA through the Space Telescope Science Institute, which is
operated by AURA, Inc., under NASA contract NAS5-26555. LB acknowledges support
from NASA grant NNX16AF40G. This research made use of the SIMBAD database,
operated at CDS, Strasbourg, France.

\begin{deluxetable}{cccccccc}     %Table3
%\rotate
\tablewidth{0pt}
\tablecolumns{8}
\tablecaption{IUE SED for G191B2B}
\tablehead{
\colhead{Wavelength (\AA)} &\colhead{Net (s$^{-1}$)}
	&\colhead{Flux\tablenotemark{a}} &\colhead{Bkg (s$^{-1}$)}
	&\colhead{Sigma\tablenotemark{a}} &\colhead{No. Obs}
	&\colhead{Exp (s$^{-1}$)} &\colhead{RMS (\%)} }
\startdata
 1152.26  &1.869e+00  &1.632e-11  &1.594e+00  &3.365e-13 &17.0  &3562.2 &13.30 \\
 1153.93  &1.672e+00  &1.822e-11  &2.111e-01  &2.407e-13 &56.0  &8208.4 &12.16 \\
 1155.61  &1.726e+00  &1.815e-11  &2.147e-01  &2.364e-13 &56.0  &8208.4 &13.24 \\
 1157.29  &1.990e+00  &1.963e-11  &2.183e-01  &2.343e-13 &56.0  &8208.4 &12.76 \\
 1158.96  &2.135e+00  &1.950e-11  &2.218e-01  &2.220e-13 &56.0  &8208.4 &14.72 \\
 1160.64  &2.264e+00  &1.882e-11  &2.253e-01  &2.084e-13 &56.0  &8208.4 & 9.68 \\
\tablenotetext{a}{erg s$^{-1}$ cm$^{-2}$ \AA$^{-1}$}
\tablecomments{Table 1 is published in its entirety in the machine-readable
      format. A portion is shown here for guidance regarding its form and
      content.}
\enddata
\label{table:g191}
\end{deluxetable}

\begin{deluxetable}{cccccccc}     %Table4
\tablewidth{0pt}
\tablecolumns{8}
\tablecaption{IUE SED for GD153}
\tablehead{
\colhead{Wavelength (\AA)} &\colhead{Net (s$^{-1}$)}
	&\colhead{Flux\tablenotemark{a}} &\colhead{Bkg (s$^{-1}$)}
	&\colhead{Sigma\tablenotemark{a}} &\colhead{No. Obs}
	&\colhead{Exp (s$^{-1}$)} &\colhead{RMS (\%)} }
\startdata
 1152.26  &3.454e-01  &2.946e-12  &1.932e-01  &1.617e-13   & 3.0   & 2399.0 &18.47 \\
 1153.93  &2.943e-01  &3.301e-12 -&2.176e-03  &8.927e-14   &10.0   &10196.1 & 9.75 \\
 1155.61  &3.165e-01  &3.370e-12 -&1.650e-03  &8.762e-14   &10.0   &10196.1 &12.24 \\
 1157.29  &3.603e-01  &3.598e-12 -&1.125e-03  &8.707e-14   &10.0   &10196.1 &10.05 \\
 1158.96  &3.866e-01  &3.543e-12 -&6.004e-04  &8.336e-14   &10.0   &10196.1 & 8.28 \\
 1160.64  &3.953e-01  &3.352e-12 -&7.707e-05  &8.060e-14   &10.0   &10196.1 &10.27 \\
\tablenotetext{a}{erg s$^{-1}$ cm$^{-2}$ \AA$^{-1}$}
\tablecomments{Table 1 is published in its entirety in the machine-readable
      format. A portion is shown here for guidance regarding its form and
      content.}
\enddata
\label{table:gd153}
\end{deluxetable}

\begin{deluxetable}{cccccccc}     %Table5
\tablewidth{0pt}
\tablecolumns{8}
\tablecaption{IUE SED for GD71}
\tablehead{
\colhead{Wavelength (\AA)} &\colhead{Net (s$^{-1}$)}
	&\colhead{Flux\tablenotemark{a}} &\colhead{Bkg (s$^{-1}$)}
	&\colhead{Sigma\tablenotemark{a}} &\colhead{No. Obs}
	&\colhead{Exp (s$^{-1}$)} &\colhead{RMS (\%)} }
\startdata
 1152.27  &3.000e-01  &3.272e-12 &-2.938e-02  &1.243e-13  & 9.0    &5781.4 &19.38 \\
 1153.94  &3.518e-01  &3.754e-12 &-3.809e-02  &1.216e-13  &10.0    &6074.9 &12.48 \\
 1155.62  &3.696e-01  &3.677e-12 &-3.746e-02  &1.168e-13  &10.0    &6074.9 & 5.90 \\
 1157.29  &4.070e-01  &3.817e-12 &-3.684e-02  &1.143e-13  &10.0    &6074.9 & 8.98 \\
 1158.97  &4.721e-01  &4.082e-12 &-3.622e-02  &1.121e-13  &10.0    &6074.9 & 7.68 \\
 1160.65  &4.802e-01  &3.813e-12 &-3.559e-02  &1.069e-13  &10.0    &6074.9 & 6.07 \\
\tablenotetext{a}{erg s$^{-1}$ cm$^{-2}$ \AA$^{-1}$}
\tablecomments{Table 1 is published in its entirety in the machine-readable
      format. A portion is shown here for guidance regarding its form and
      content.}
\enddata
\label{table:gd71}
\end{deluxetable}

\begin{deluxetable}{cccccccc}     %Table6
\tablewidth{0pt}
\tablecolumns{8}
\tablecaption{IUE SED for GRW+70$^{\circ}$5824}
\tablehead{
\colhead{Wavelength (\AA)} &\colhead{Net (s$^{-1}$)}
	&\colhead{Flux\tablenotemark{a}} &\colhead{Bkg (s$^{-1}$)}
	&\colhead{Sigma\tablenotemark{a}} &\colhead{No. Obs}
	&\colhead{Exp (s$^{-1}$)} &\colhead{RMS (\%)} }
\startdata
 1152.26  &1.439e-01  &1.369e-12  &5.961e-02  &6.907e-14  & 7.0  & 7413.7 &11.99 \\
 1153.93  &1.435e-01  &1.348e-12  &4.069e-02  &5.633e-14  &10.0  &10580.7 & 7.38 \\
 1155.61  &1.661e-01  &1.432e-12  &4.083e-02  &5.547e-14  &10.0  &10580.7 &12.71 \\
 1157.29  &1.872e-01  &1.526e-12  &4.097e-02  &5.352e-14  &10.0  &10580.7 & 9.74 \\
 1158.96  &1.947e-01  &1.465e-12  &4.111e-02  &5.033e-14  &10.0  &10580.7 & 5.37 \\
 1160.64  &2.005e-01  &1.384e-12  &4.124e-02  &4.672e-14  &10.0  &10580.7 & 7.86 \\
\tablenotetext{a}{erg s$^{-1}$ cm$^{-2}$ \AA$^{-1}$}
\tablecomments{Table 1 is published in its entirety in the machine-readable
      format. A portion is shown here for guidance regarding its form and
      content.}
\enddata
\label{table:grw}
\end{deluxetable}

\begin{deluxetable}{cccccccc}     %Table7
\tablewidth{0pt}
\tablecolumns{8}
\tablecaption{IUE SED for HZ21}
\tablehead{
\colhead{Wavelength (\AA)} &\colhead{Net (s$^{-1}$)}
	&\colhead{Flux\tablenotemark{a}} &\colhead{Bkg (s$^{-1}$)}
	&\colhead{Sigma\tablenotemark{a}} &\colhead{No. Obs}
	&\colhead{Exp (s$^{-1}$)} &\colhead{RMS (\%)} }
\startdata
 1152.26  &1.060e-01  &9.647e-13  &5.584e-02  &2.857e-14    &14.0 &28312.8 &17.37 \\
 1153.93  &1.150e-01  &9.888e-13  &5.356e-02  &2.520e-14    &16.0 &34072.3 &14.81 \\
 1155.61  &1.247e-01  &1.006e-12  &5.369e-02  &2.439e-14    &16.0 &34072.3 &12.47 \\
 1157.29  &1.394e-01  &1.046e-12  &5.382e-02  &2.357e-14    &16.0 &34072.3 &19.28 \\
 1158.96  &1.426e-01  &9.967e-13  &5.260e-02  &2.188e-14    &17.0 &35332.1 &13.27 \\
 1160.64  &1.516e-01  &9.709e-13  &5.273e-02  &2.056e-14    &17.0 &35332.1 &13.20 \\
\tablenotetext{a}{erg s$^{-1}$ cm$^{-2}$ \AA$^{-1}$}
\tablecomments{Table 1 is published in its entirety in the machine-readable
      format. A portion is shown here for guidance regarding its form and
      content.}
\enddata
\label{table:hz21}
\end{deluxetable}

\begin{deluxetable}{cccccccc}     %Table8
\tablewidth{0pt}
\tablecolumns{8}
\tablecaption{IUE SED for HZ43}
\tablehead{
\colhead{Wavelength (\AA)} &\colhead{Net (s$^{-1}$)}
	&\colhead{Flux\tablenotemark{a}} &\colhead{Bkg (s$^{-1}$)}
	&\colhead{Sigma\tablenotemark{a}} &\colhead{No. Obs}
	&\colhead{Exp (s$^{-1}$)} &\colhead{RMS (\%)} }
\startdata
 1152.26  &6.090e-01  &5.383e-12  &3.758e-01  &1.635e-13 &10.0 &4646.9 &15.53 \\
 1153.93  &7.200e-01  &5.756e-12  &5.555e-01  &1.532e-13 &12.0 &5308.7 &17.83 \\
 1155.61  &7.274e-01  &5.510e-12  &5.559e-01  &1.470e-13 &12.0 &5308.7 &14.41 \\
 1157.29  &7.897e-01  &5.528e-12  &5.562e-01  &1.470e-13 &12.0 &5308.7 &16.23 \\
 1158.96  &8.672e-01  &5.614e-12  &5.566e-01  &1.388e-13 &12.0 &5308.7 &10.60 \\
 1160.64  &9.610e-01  &5.716e-12  &5.569e-01  &1.342e-13 &12.0 &5308.7 &12.88 \\
\tablenotetext{a}{erg s$^{-1}$ cm$^{-2}$ \AA$^{-1}$}
\tablecomments{Table 1 is published in its entirety in the machine-readable
      format. A portion is shown here for guidance regarding its form and
      content.}
\enddata
\label{table:hz43}
\end{deluxetable}

\newpage

\bibliographystyle{apj}
\bibliography{../../pub/paper-bibliog}

\end{document}